\documentclass[prc,unsortedaddress,groupedaddress,preprint,amsmath,amsfonts,amssymb,showpacs,floatfix]{revtex4-1}

\usepackage{graphicx}

\usepackage{amsmath}
\usepackage{bm}
\usepackage{color}

\begin{document}

\title{Primordial ${\alpha} + d \to {}^{6}{\rm Li} + \gamma$  reaction and second Lithium puzzle }

\author{A. M. Mukhamedzhanov}
\email{akram@comp.tamu.edu}
\affiliation{Cyclotron Institute, Texas A\&M University, College Station, TX 77843, USA }
\author{Shubhchintak}
\email{Shub.Shubhchintak@tamuc.edu}
\affiliation{Department of Physics, Texas A\&M University-Commerce, Commerce, TX 75429, USA  }
\author{C. A.  Bertulani}
\email{carlos.bertulani@tamuc.edu}
\affiliation{Department of Physics, Texas A\&M University-Commerce, Commerce, TX 75429, USA }
\affiliation{Department of Physics and Astronomy, Texas A\&M University, College Station, Texas 77843, USA}

\date{Today}

\begin{abstract}
During the Big Bang,   ${}^{6}{\rm Li}$  was synthesized via the ${}^{2}{\rm H}(\alpha,\gamma){}^{6}{\rm Li}$ reaction. 
After almost 25 years of the failed attempts to measure  the ${}^{2}{\rm H}(\alpha,\gamma){}^{6}{\rm Li}$  reaction in the lab at the Big Bang energies,  just recently the LUNA collaboration  presented the first successful  measurements at two different Big Bang energies  [M. Anders {\it et al.}, Phys. Rev. Lett. {\bf 113}, 042501 (2014)]. In this paper we will discuss how to improve the accuracy of the direct experiment. To this end the photon's angular distribution is calculated in the potential model. It contains contributions from  electric dipole and quadrupole transitions and their interference, which dramatically  changes the  photon's angular distribution. The calculated distributions at different Big Bang energies have a single peak  at $\sim 50^{\circ}$. These calculations provide the best kinematic conditions to measure the ${}^{2}{\rm H}(\alpha,\gamma){}^{6}{\rm Li}$ reaction. The expressions for the total cross section and astrophysical factor are also derived by integrating the differential cross section over the photon's solid angle. The LUNA  data are in excellent agreement with our calculations using a potential approach combined with a well established  asymptotic normalization coefficient for ${}^{6}{\rm Li} \to \alpha +d$. Comparisons of the available experimental data for the $S_{24}$ astrophysical factor  and different calculations are presented.  The  Big Bang lithium isotopic ratio ${}^{6}{\rm Li}/^{7}{\rm Li} = (1.5 \pm 0.3)\times 10^{-5}$ following from the LUNA data and the present analysis are discussed in the context of the disagreement between the observational data and the standard Big Bang model,  which constitutes the second Lithium problem.

\end{abstract}

\maketitle

\section{Introduction}
The primordial nuclei were formed during the first 20 minutes after the Big Bang.
Among them the lithium isotopes, ${}^{7}{\rm Li}$ and a much smaller amount of ${}^{6}{\rm Li}$, were synthesized.  
Later on, cosmic rays,  novae and pulsations of  AGB stars were the main generators of the ${}^{7}{\rm Li}$ isotope, and  ${}^{6}{\rm Li}$ was formed mainly by cosmic rays.  In 1982, two important papers \cite{spitea, spiteb} for the first time  noted  that metal-poor ($-2.4\leq [{\rm Fe/H}] \leq-1.4$), warm ($5700 \leq T_{eff}  \leq 6250$ K)  dwarf stars demonstrated remarkably constant ${}^{7}{\rm Li}$ abundance (Spite plateau), which does not depend on metallicity and effective temperature.  It was quite a surprising observation because depletion of lithium over such a broad temperature range should be significant.  Because it was impossible to explain the existence of  the Spite plateau over a wide range of temperatures, it was suggested  that  no depletion of  ${}^{7}{\rm Li}$ took place in the observed dwarf warm stars and that the  constant abundance of ${}^{7}{\rm Li}$ is the primordial one.  However, this interpretation of the Spite plateau was periodically challenged.  For example, in \cite{aoki,sbordone}  the meltdown of the Spite plateau was discovered in some low metallicity stars.

Explanation of the lithium abundance in low metallicity stars in the halo of our Galaxy where the lithium abundance demonstrates independence on metallicity, forming the Spite plateau,  became one of the hot  topics in modern cosmology/nuclear astrophysics. Note that the observations  of the primordial lithium are restricted to white dwarfs because the loosely bound ${}^{7}{\rm Li}$ nuclei are easily destroyed by the ${}^{7}{\rm Li}({\rm p},\alpha){}^{4}{\rm He}$ reaction when the temperature exceeds $2.6\times10^{6}$ K.  For this reason, red giants cannot be used to determine the lithium primordial abundance. 

In the standard Big Bang nucleosynthesis  model, ${}^{7}{\rm Li}$ is formed right after the Big Bang, together with ${}^{1}{\rm H},\, {}^{2}{\rm H},\,{}^{3}{\rm He}$ and ${}^{4}{\rm He}$.  The primordial reactions start from the deuteron formation ${\rm p +n \to d} + \gamma$. The deuteron's yield  depends on the primordial  baryon/photon ratio   $\eta_{B}$.  Because the deuterons are seeds, which are necessary to synthesize heavier elements,  the abundance of heavier elements, and lithium in particular, also depends on $\eta_{B}$.   $\,{}^{2}{\rm H}\,$ and $\,{}^{7}{\rm Li}\,$  are  two primordial nuclei which is most sensitive to $\eta_{B}$.  

The abundance of the primordial ${}^{7}{\rm Li}$ within the framework of the standard Big Bang scenario, calculated using the extended reaction network
and nine years of WMAP results \cite{hinshaw},  is ${}^{7}{\rm Li/H} = 5.13 \times 10^{-10}$.  It is  ${}^{7}{\rm Li/H} = (4.56- 5.34) \times 10^{-10}$  \cite{coc2014} based on the Planck results  \cite{planck} plus the information about the lensing potential  and  ground-based high resolution experiments. The latter is considered to be the most up-to-date estimation of the  ${}^{7}{\rm Li}$ isotope abundance within the  standard Bing Bang scenario. This abundance  remains significantly higher than  more recent  observations in metal poor halo stars \cite{sbordone}: ${}^{7}{\rm Li/H} = 1.58^{+0.35}_{ - 0.28} \times  10^{-10}$. The  shortage  of the observed  ${}^{7}{\rm Li}$ compared to the standard Big Bang predictions  represents the so-called first Lithium puzzle.  

During the Big Bang  a very small amount of ${}^{6}{\rm Li}$  was synthesized via the ${}^{2}{\rm H}(\alpha,\gamma){}^{6}{\rm Li}$ reaction.
 Later,  ${}^{6}{\rm Li}$ was mostly formed by  cosmic rays.  The primordial ${}^{6}{\rm Li}$ is assumed to be present in the gas from which the stars were formed.  Unlike most of the other elements, when ${}^{6}{\rm Li}$ is synthesized inside the stars by hydrostatic
nucleosynthesis, it is quickly  destroyed.  But in the atmosphere of  the halo metal-poor warm dwarfs,  the primordial ${}^{6}{\rm Li}$ can survive for 13 billion years
not being affected by  cosmic rays, although its survival can be questioned. ${}^{7}{\rm Li}$ is used to help determine the primordial Big Bang ${}^{6}{\rm Li}$ 
abundance. First, the presence of ${}^{6}{\rm Li}$ constrains the destruction of ${}^{7}{\rm Li}$, because ${}^{6}{\rm Li}$ is more easily destroyed than ${}^{7}{\rm Li}$. Besides, if ${}^{6}{\rm Li}$ was formed before the formation of the stars, then the same is true for ${}^{7}{\rm Li}$.  

Stellar ${}^{7}{\rm Li}$ abundance is usually determined from the resonance line at 670.8 nm but in exceptional cases also from the weaker line at 610.4 nm. The isotope ${}^{6}{\rm Li}$  can be detected through the isotopic shift in the Li I 670.8 nm line. The distortion of the line profile is very small and therefore requires  very high quality  spectra. Ref. \cite{asplund2006}  reported for the first time the detection of a high abundance of  ${}^{6}{\rm Li}$ in  very metal-poor stars. The authors concluded that the observed ${}^{6}{\rm Li}$  was formed during  Big-Bang nucleosynthesis. The detection of ${}^{6}{\rm Li}$ was based on the fact noted above, that the presence of ${}^{6}{\rm Li}$ in the stellar atmosphere  causes an asymmetry in the Li 670.8 nm line. The average ${}^{6}{\rm Li}/{}^{7}{\rm Li}$ isotopic ratio in the nine stars, in which ${}^{6}{\rm Li}$ was detected, was  ${}^{6}{\rm Li}/{}^{7}{\rm Li}  \sim 5 \times 10^{-2}$ \cite{asplund2006}. Such a high isotopic ratio of the primordial lithium isotopes in the metal-poor stars contradicts  the Big-Bang based model predictions  ${}^{6}{\rm Li}/{}^{7}{\rm Li}  \sim  10^{-5}$ \cite{coc2014} and cannot be explained by the galactic cosmic rays.  This  disagreement  between the observations and Big Bang predictions  of the lithium isotopic ratio constitutes the second Lithium problem.  

Later it was pointed out in \cite{cayrel2007} that the  line asymmetry caused by convection in the photospheres of metal-poor  stars is practically indistinguishable from the asymmetry produced by  a weak ${}^{6}{\rm Li}$ distortion of a  symmetric ${}^{7}{\rm Li}$ line. Hence, the ${}^{6}{\rm Li}$ abundance obtained in  \cite{asplund2006} could be significantly overestimated,  and the result obtained in \cite{asplund2006}  can be considered only as an upper limit of the lithium isotopic ratio.  In Ref. \cite{lind} the lithium isotopic ratio was reanalyzed within the framework of the 3D, non-local thermodynamic equilibrium (NLTE) model.  The authors came to the conclusion that ``'the observational support for a significant and non-standard ${}^{6}{\rm Li}$  production source in the early universe is substantially weakened by our findings" \cite{lind},  which opens a way to a hope that the primordial abundance of ${}^{6}{\rm Li}$ calculated in the standard Big Bang  nucleosynthesis can be eventually reconciled with observational data.
 
The yields of the observed and predicted primordial ${}^{7}{\rm Li}$ are established quite well \cite{coc2014}.
If the observed ${}^{6}{\rm Li}$  is  primordial (a Big Bang product) then its  abundance is determined by the ${}^{2}{\rm H}(\alpha,\gamma){}^{6}{\rm Li}$  reaction.  The first successful attempt to measure the ${}^{2}{\rm H}(\alpha,\gamma){}^{6}{\rm Li}$ reaction was reported in \cite{robertson}
where residual ${}^{6}{\rm Li}$ nuclei were detected. The  astrophysical factor was measured in the vicinity of the first resonance  ${}^{6}{\rm Li}(3^{+})$  at the relative $\alpha{\rm -d}$ energy $E=0.712$ MeV and at higher energies. But no data were obtained at  Big Bang energies, $30 \lesssim E \lesssim 400$ keV.
 In \cite{mohr}  the astrophysical  $S_{24}(E)$  factor was also measured only at the resonance energy, using  in-beam spectroscopy.  

In Ref. \cite{kiener} for the first time, an attempt was made to measure the astrophysical factor at the Big Bang energies, using the Coulomb breakup of ${}^{6}{\rm Li}$  at $26$ MeV/A energy on a ${}^{208}{\rm Pb}$ target.  However, only an upper limit was established.  The failure of this indirect attempt to measure the $S_{24}(E)$  astrophysical factor could be anticipated because the $E1$ transition, which usually dominates, is suppressed in the case under consideration:  the effective charge for the dipole transition is very  small owing to practically the same charge/mass ratio for $\alpha$-particle and deuteron. Because the Coulomb dissociation cross section is dominated by the $E2$ transition, the obtained data may be considered only as an upper  limit.  
After that, another unsuccessful attempt to measure  the $S_{24}(E)$ factor ended  with an upper limit  $S_{24}(53\, {\rm keV})  <  2.0 \times 10^{-7}$ MeVb and a pessimistic conclusion that it would be impossible to measure directly $S_{24}(E)$ at Big Bang energies \cite{ceccil}. 

The second attempt to use the indirect Coulomb dissociation technique was made in \cite{hammache}, where the  breakup of ${}^{6}{\rm Li}$ ions at  $150$ MeV/A on a ${}^{208}{\rm Pb}$ target was measured.   However, in this case,  the breakup was dominated by  nuclear breakup, which overwhelmed the Coulomb breakup.  Hence,  no information about  $S_{24}(E)$ was extracted from the analysis of the breakup data.  Further, in Ref. \cite{hammache} the astrophysical factor was calculated  using a two-body potential model (see below).  
Finally, after almost 25 years of  failed attempts to measure the ${}^{2}{\rm H}(\alpha,\gamma){}^{6}{\rm Li}$  reaction at the Big Bang energies,   just recently the LUNA collaboration  presented the first successful  measurements at two different Big Bang energies \cite{LUNA}.  Definitely it is a remarkable achievement  in the studies of  Big Bang nucleosynthesis. 

In this work we discuss the  astrophysical ${}^{2}{\rm H}(\alpha, \gamma){}^{6}{\rm Li}$ reaction within the framework of the potential approach and  impact on  experimental measurements.  For the first time, we present  the angular distribution of the photons emitted in this direct radiative capture. Although the photon differential cross section is being derived for the  ${}^{2}{\rm H}(\alpha, \gamma){}^{6}{\rm Li}$ process, it can be applied for any direct electric radiative capture reaction. 
The calculated angular distributions provide the best kinematics to be used in the measurement of the emitted photons, which differ from the one used in the LUNA experiment.  Optimal kinematics will allow one to  decrease significantly the uncertainty of direct measurements of the ${}^{2}{\rm H}(\alpha, \gamma){}^{6}{\rm Li}$ process compared to the uncertainties in the LUNA experiment. By integrating the differential cross section over the photon solid angle, the total cross section and astrophysical factor of the direct radiative capture are derived. The calculations of the photon's angular distribution and astrophysical $S_{24}(E)$ factor are done in the potential model using the well determined asymptotic normalization coefficient for the virtual decay ${}^{6}{\rm Li} \to \alpha +d$. The  primordial ${}^{6}{\rm Li}$ abundance is presented. 

\section{Photon differential cross sections, total cross sections and astrophysical S-factors}

\subsection{ Photon angular distribution in direct radiative capture }

In this section the expression for the  angular distribution of the photons emitted in the $\alpha({\rm d},\gamma){}^{6}{\rm Li}$ direct radiative capture  is derived and further  simplified  in the subsequent section.
This result can help  to  improve future experiments on this reaction by decreasing their uncertainties.
Often photon angular distributions are not discussed in the papers dealing with  measurements of the astrophysical factors. That is why we believe it is timely to do it.
Besides by integrating the photon differential cross section over the photon's solid angle the total cross section and the astrophysical factor can be derived. 

We  consider the photon angular distribution  taking into account the spin-orbit interaction in the initial state. Hence,  the initial scattering wave function depends  on the initial $\alpha-{\rm d}$
relative orbital angular momentum  $l_{i}$,  the channel spin $s$ and the total angular momentum in the initial channel $J_{i}$. In the case under consideration $s=J_{d}$, where $J_{d}=1$  is the spin of the deuteron.
The differential cross section of the emitted photons with momentum ${\rm {\bf k}}_{\gamma}$ and helicity $\lambda=\pm 1$ in 
 the electromagnetic  transition from the initial continuum state $l_{i},\,s,\,J_{i}$ to the final state  $l_{f}\,,s,\,J_{f}$ in the center-of-mass of ${}^{6}{\rm Li}$  is given by
\begin{align}
\frac{{\rm d}\sigma_{\lambda}}{{\rm d}\Omega}  \sim k_{\gamma}^{2}\, \left|-\frac{1}{c}\int\,{\rm d}{\rm {\bf r}}\,
\left<\varphi_{{}^{6}{\rm Li}}(\zeta_{\alpha},\,\zeta_{d};{\rm {\bf r}}_{\alpha\,d})\,\big|{\rm {\bf {\hat J}}}({\rm {\bf r}}) \big| \Psi_{{\rm {\bf k}}}^{(+)}(\zeta_{\alpha},\,\zeta_{d};{\rm {\bf r}}_{\alpha\,d})\right>\cdot {\rm {\bf A}}^{*}_{\lambda\,{\rm {\bf k}}_{\gamma}}({\rm {\bf r}}) \right|^{2}.
\label{diffcrsect1}
\end{align}
Here, ${\rm {\bf A}}_{\lambda\,{\rm {\bf k}}_{\gamma}}({\rm {\bf r}})$  is the vector-potential of the photon with helicity $\lambda$ and momentum ${\rm {\bf k}}_{\gamma}$ at  coordinate ${\rm {\bf r}}$.
The initial  wave function is
\begin{align}
\Psi_{{\rm {\bf k}}}^{(+)}(\zeta_{\alpha},\,\zeta_{d};\,{\rm {\bf r}}_{\alpha\,d})= \varphi_{\alpha}\,(\zeta_{\alpha})\,\varphi_{d}(\zeta_{d})\,\psi^{(+)}({\rm {\bf k}},\,{\rm{\bf r}}_{\alpha\,d}),
\label{Psiin1}
\end{align}
$\varphi(\zeta_{i})$ is the bound-state wave function of  nucleus $i$ with the set of the internal coordinates $\zeta_{i}$, which includes spin-isospin variables.  $\psi^{(+)}({\rm {\bf k}},\,{\rm {\bf r}}_{\alpha\,d})$ is the $\alpha-{\rm d}$ scattering wave function in the initial state, $ \,{\rm {\bf r}}_{\alpha\,d}$ is the radius-vector connecting the centers of mass of the $\alpha$-particle and the deuteron,  $\,\,{\rm {\bf k}}$ is the initial $\alpha-{\rm d}$ relative momentum related to the initial relative kinetic energy as $E=k^{2}/(2\,\mu_{\alpha\,{\rm d}})$,  where $\mu_{\alpha\,{\rm d}}$ is the  $\,\alpha-{\rm d}$ reduced mass. The momentum of the emitted photon is $k_{\gamma}= (E+\varepsilon)/\hbar$  and expressed in fm${}^{-1}$, $\varepsilon$ is the binding energy for the virtual decay ${}^{6}{\rm Li} \to \alpha + {\rm d}$.
The antisymmetrization between the nucleons of the $\alpha$-particle and the deuteron is neglected.  Note that all the kinematic factors defining the photon differential cross section including the spin-dependent factors will be recovered later.

We use the long wavelength approximation, which is valid for $k_{\gamma}\,R_{\alpha\,d} <<1$.  Here  $ R_{\alpha\,{\rm d}}\,$ is the effective $\alpha-{\rm d}$ distance determined so that 
distances $r \sim R_{\alpha\,{\rm d}}\,$ give the dominant contribution to the amplitude of the direct radiative capture. The long electromagnetic  wavelength  of the emitted radiation  allows us to approximate the charge current density by the current density of the point-like $\alpha$-particle and deuteron neglecting their internal structure:
\begin{align}
{\rm {\bf {\hat J}}}({\rm {\bf r}}) = \frac{Z_{d}\,e}{2\,m_{d}}\,\big[\delta({\rm {\bf r}}- {\rm {\bf r}}_{d})\,{\rm {\bf {\hat p}}}_{d} + {\rm {\bf {\hat p}}}_{d}\,\delta({\rm {\bf r}} - {\rm {\bf r}}_{d})\big]  + \frac{Z_{\alpha}\,e}{2\,m_{\alpha}}\,\big[\delta({\rm {\bf r}} - {\rm {\bf r}}_{\alpha})\,{\rm {\bf {\hat p}}}_{\alpha}  + {\rm {\bf {\hat p}}}_{\alpha}\,\delta({\rm {\bf r}}- {\rm{\bf r}}_{\alpha})\big],
\label{chargedensoper1}
\end{align}
where ${\rm {\bf {\hat p}}}_{i}= -i\,\hbar\,{\partial  }/{ {\partial }{\rm{\bf r}}_{i} }$ is the momentum  operator,  ${\rm {\bf r}}_{d}= 
- (m_{\alpha}/m_{\alpha\,d})\,{\rm {\bf r}}_{\alpha\,d}$  and ${\rm {\bf r}}_{\alpha}=(m_{d}/m_{\alpha\,d})\,{\rm {\bf r}}_{\alpha\,d}$ are the coordinates of the centers of mass of the deuteron and alpha-particle, respectively, $m_{i}$ and  $Z_{i}$  are the mass and atomic number  of nucleus $i$ and $m_{ij}= m_{i} + m_{j}$. 
We neglect here the spin contribution to the current density because below we consider only the electric transitions which are largely due to the charge current. 

Now  the overlap function of the bound-state wave functions of ${}^{6}{\rm Li},\,\alpha$-particle and deuteron can be introduced:
\begin{align}
&I_{l_{f}\,s\,J_{f}}({\rm {\bf r}}_{\alpha\,d})= \left<\varphi_{\alpha}(\zeta_{\alpha})\,\varphi_{d}(\zeta_{d})\big|\varphi_{{}^{6}{\rm Li}}(\zeta_{\alpha},\,\zeta_{d}; {\rm {\bf r}}_{\alpha\,d})\right>  \nonumber\\
&= \sum\limits_{m_{l_{f}}\,m_{s}''}\,\left<l_{f}\,m_{l_{f}}\,\,s\, m_{s}''|J_{f}\,M_{f}\right>\,Y_{l_{f}m_{l_{f}}}({\rm {\bf {\hat r}}}_{\alpha\,d})\,\chi_{s\,m_{s}'' }\,I_{l_{f}\,s\,J_{f}}(r_{\alpha\,d}),
\label{overlapfunction1}
\end{align}
where $\,I_{l_{f}\,s\,J_{f}}(r_{\alpha\,d})$ is the radial overlap function,  $\left<l_{f}\,m_{l_{f}}\,s\,m_{s}''|J_{f}\,M_{f}\right>$  is the Clebsch-Gordan coefficient,
 $\,l_{f}$ is the $\alpha-{\rm d}$ relative orbital angular momentum in the bound state, $\,J_{f}=1$ is the spin of the ground state of ${}^{6}{\rm Li}$;
$\chi_{s\,m_{s}''}$ is the spin wave function describing the state with the channel spin $s$ and its projection $m_{s}''$,
${\rm {\bf {\hat r}}} =  {\rm {\bf r}}/{\rm r}$ is the unit vector.
The integration in the matrix element $\left<\varphi_{\alpha}(\zeta{\alpha})\,\varphi_{d}(\zeta_{d})\big|\varphi_{{}^{6}{\rm Li}}(\zeta_{\alpha},\,\zeta_{d}; {\rm {\bf r}}_{\alpha\,d})\right>$ is taken over all the internal coordinates $\zeta_{\alpha}$ and $\zeta_{d}$  making  the overlap function depending only on the radius-vector ${\rm {\bf r}}_{\alpha\,d}$.

In the  peripheral region the radial overlap function  is given by
\begin{align}
I_{l_{f}sJ_{f}}(r_{\alpha\,d}) \stackrel{r_{\alpha\,d}> r_{0}}{\approx} C_{l_{f}sJ_{f}}\,W_{-\eta,\,l_{f}+1/2}(2\,\kappa\,r_{\alpha\,d})/r_{\alpha\,d},    
\label{radovas1}
\end{align}
where $C_{l_{f}sJ_{f}}$ is the asymptotic normalization coefficient (ANC)  for the virtual decay ${}^{6}{\rm Li} \to \alpha + {\rm d}$ expressed in fm$^{-1/2}$,    $\,W_{-\eta_{f},\,l_{f}+1/2}(2\,\kappa\,r_{\alpha\,d})$ is the Whittaker function determining the radial shape of the overlap function beyond of the $\alpha-d$ nuclear interaction region, $\eta_{f}= ({Z_{\alpha}\,Z_{d}\,e^{2}}/{\hbar\,c})({\mu_{\alpha\,d}\,c}/{\hbar})({1}/{\kappa})$ is the  Coulomb  $\alpha-{\rm d}$ bound-state parameter, $\,\kappa= {\sqrt{2\,\mu_{\alpha\,d}\,c^{2}\,\varepsilon}}/{\hbar\,c}$ is the $\alpha-d$ bound-state wave number expressed in fm$^{-1}$.
The radial overlap function is expressed in fm$^{-3/2}$.  $\,r_{0}$ is the channel radius, which is selected so that at $r_{\alpha\,d} > r_{0}$ the nuclear interaction between the deuteron and $\alpha$-particle is negligible. 

The matrix element  in (\ref{diffcrsect1}) now can be rewritten as
\begin{align}
&\frac{1}{c}\int\,{\rm d}{\rm {\bf r}}\,
\left<\varphi_{{}^{6}{\rm Li}}(\zeta_{\alpha},\,\zeta_{d};{\rm {\bf r}}_{\alpha\,d})\,\big|{\rm {\bf {\hat J}}}({\rm {\bf r}}) \big| \Psi^{(+)}(\zeta_{\alpha},\,\zeta_{d};{\rm {\bf r}}_{\alpha\,d})\right>\cdot{\rm {\bf A}}^{*}_{\lambda\,{\rm {\bf k}}_{\gamma}}({\rm {\bf r}}) \nonumber\\
&=\frac{1}{c}\int\,{\rm d}{\rm {\bf r}}\,
\left<I_{l_{f}\,s\,J_{f}}({\rm {\bf r}}_{\alpha\,d})\,\big|{\rm {\bf {\hat J}}}({\rm {\bf r}}) \big| \psi^{(+)}({\rm {\bf k}},\,{\rm{\bf r}}_{\alpha\,d})\right>\cdot {\rm {\bf A}}^{*}_{\lambda\,{\rm {\bf k}}_{\gamma}}({\rm {\bf r}}).
\label{trmatrelem1}
\end{align}

To simplify further  this matrix element we need to use the multipole expansion of the vector potential \cite{balashov,eisenberg} :
\begin{align}
&{\rm {\bf A}}_{\lambda\,{\rm {\bf k}}_{\gamma}}({\rm {\bf r}}) = \frac{1}{2\,\pi}\,\sqrt{\frac{\hbar\,c}{k_{\gamma}}}\,{\rm {\bf e}}_{\lambda\,{\rm {\bf k}}_{\gamma}}\, e^{i\,{\rm {\bf k} _{\gamma}} \cdot {\rm {\bf r}} }  
= \frac{1}{2\,\sqrt{2\,\pi}\,k_{\gamma}}\,\sum\limits_{L\,M}\,\sqrt{2\,L+1}\,D_{M\,\lambda}^{L}(\varphi, \theta,0)  \nonumber\\
&\times \Big({\rm {\bf A}}_{ek_{\gamma}LM}({\rm {\bf r}}) +\lambda\,{\rm {\bf A}}_{mk_{\gamma}LM}({\rm {\bf r}})\Big).
\label{multexpA1}
\end{align}
Here, ${\rm {\bf e}}_{\lambda\,{\rm {\bf k}}_{\gamma}}$ is the unit  polarization vector of the plane wave, which is orthogonal to the photon momentum ${\rm {\bf k}}_{\gamma}$,  $\,\,{\rm {\bf A}}_{ek_{\gamma}LM}({\rm {\bf r}})$ and ${\rm {\bf A}}_{mk_{\gamma}LM}({\rm {\bf r}})$ are the eletric and magnetic multipoles, correspondingly. 
In the system $z \| {\rm {\bf k}}_{\gamma}$ the helicity of the circularly polarized photon  $ \lambda= \pm 1$.   $D_{M\,\lambda}^{L}(\varphi,\,\theta,0)$ is the Wigner $D$-function, $L$ is the multipolarity of the transition. 
 In Eq. (\ref{multexpA1})  only the electric multipoles ${\rm {\bf A}}_{e\,k_{\gamma}\,L\,M}({\rm {\bf r}})$  will be taken into account because for the reaction under consideration the contribution of the magnetic multipoles ${\rm {\bf A}}_{m\,k_{\gamma}\,L\,M}({\rm {\bf r}})$  is negligible \cite{nollett2001}.
Following Ref. \cite{balashov},    ${\rm {\bf A}}_{e\,k_{\gamma}\,L\,M}({\rm {\bf r}})$  can be rewritten as
\begin{align}
&{\rm {\bf A}}_{e\,k_{\gamma}\,L\,M}({\rm {\bf r}}) = -2\,i^{L}\,\sqrt{\frac{\hbar\,c}{k_{\gamma}}}\,\Big[{\rm {\bm \nabla}}_{\rm {\bf r}} \times \big( j_{L}(k_{\gamma}\,r)\,{\rm {\bf Y}}_{LM}^{L}({\rm {\bf{\hat r}}}) \big) \Big]               \nonumber\\
&= 2\,i^{L-1}\,\sqrt{\hbar\,c\,k_{\gamma}}\,\Big[\sqrt{\frac{L+1}{2\,L+1}}\,j_{L-1}(k_{\gamma}\,r) \,{\rm {\bf Y}}_{LM}^{L-1}({\rm {\bf {\hat r}}})  - \sqrt{\frac{L}{2\,L+1}}\,j_{L+1}(k_{\gamma}\,r)\,{\rm {\bf Y}}_{LM}^{L+1}({\rm {\bf {\hat r}}}) \Big],
\label{ElmultAekLM1}
\end{align} 
where ${\rm {\bf Y}}_{LM}^{{\tilde L}}({\rm {\bf {\hat r}}})$ is the  vector spherical harmonics \cite{balashov,eisenberg} and $j_{L}(k_{\gamma}\,r)$ is the spherical Bessel function. 

Now  the matrix element  (\ref{trmatrelem1})  can be reduced to
\begin{align}
&\frac{1}{c}\int\,{\rm d}{\rm {\bf r}}\,
\left<I_{l_{f}\,s\,J_{f}}({\rm {\bf r}}_{\alpha\,d})\,\big|{\rm {\bf {\hat J}}}({\rm {\bf r}})\,\big|\psi^{(+)}({\rm {\bf k}},\,{\rm{\bf r}}_{\alpha\,d})\right>\cdot{\rm {\bf A}}^{*}_{\lambda\,{\rm {\bf k}}_{\gamma}}({\rm {\bf r}})  \nonumber\\
&=  \sqrt{ \frac{\hbar}{2\,\pi\,c\,k_{\gamma}}}\,\sum\limits_{L\,M}\,i^{-L+1}\,\sqrt{2\,L+1}\,\big(D_{M\,\lambda}^{L}(\varphi, \theta,0)\big)^{*}\,\int {\rm d}{\rm {\bf r}}
\left<I_{l_{f}\,s\,J_{f}}({\rm {\bf r}}_{\alpha\,d})\big|{\rm {\bf {\hat J}}}({\rm {\bf r}}) \big| \psi^{(+)}({\rm {\bf k}},\,{\rm{\bf r}}_{\alpha\,d})\right>                                                                            \nonumber\\
&\times \Big[\sqrt{\frac{L+1}{2\,L+1}}\,j_{L-1}(k_{\gamma}\,r) \,({\rm {\bf Y}}_{LM}^{L-1}({\rm {\bf {\hat r}}}))^{*}      
 - \sqrt{\frac{L}{2\,L+1}}\,j_{L+1}(k_{\gamma}\,r)\,({\rm {\bf Y}}_{LM}^{L+1}({\rm {\bf {\hat r}}}))^{*} \Big]       \nonumber\\
& \approx  \sqrt{ \frac{\hbar}{2\,\pi\,c\,k_{\gamma}}}\,\sum\limits_{L\,M}\,\frac{i^{-L+1}\,k_{\gamma}^{L-1}}{(2\,L-1)!!}\,\sqrt{L+1}\,\big(D_{M\,\lambda}^{L}(\varphi, \theta,0)\big)^{*}\int{\rm d}{\rm {\bf r}}
\left<I_{l_{f}\,s\,J_{f}}({\rm {\bf r}}_{\alpha\,d})\big|{\rm {\bf {\hat J}}}({\rm {\bf r}}) \big| \psi^{(+)}({\rm {\bf k}},\,{\rm{\bf r}}_{\alpha\,d})\right>                                                                            \nonumber\\
&\times r^{L-1} \,\big({\rm {\bf Y}}_{LM}^{L-1}({\rm {\bf {\hat r}}})\big)^{*}.
\label{matrelementlwfl1}
\end{align}
In the long wavelength approximation $k_{\gamma}\,r <<1$, $\,\,j_{L}(k_{\gamma}\,r) \approx (k_{\gamma}\,r)^{L}/(2\,L+1)!!$. 
Hence, the lowest  partial waves dominate and  the term containing  $j_{L+1}(k_{\gamma}\,r) \approx (k_{\gamma}\,r)^{L+1}/(2\,L+3)!!$ is  small compared to the term containing $j_{L-1}(k_{\gamma}\,r) \approx (k_{\gamma}\,r)^{L-1}/(2\,L-1)!!$ and can be neglected. 

Taking into account that  \cite{edmonds}
\begin{align}
{\rm {\bm \nabla}}_{\rm {\bf r}} [r^{L}\,Y_{L\,M}({\rm {\bf {\hat r}}})]= \sqrt{L(2\,L+1)}\,r^{L-1}\,{\rm {\bf Y}}_{L\,M}^{L-1}({\rm {\bf {\hat r}}}),
\label{gradeqn1}
\end{align}
Equation (\ref{matrelementlwfl1}) can be reduced to
\begin{align}
&\frac{1}{c}\int\,{\rm d}{\rm {\bf r}}\,
\left<I_{l_{f}\,s\,J_{f}}({\rm {\bf r}}_{\alpha\,d})\big|{\rm {\bf {\hat J}}}({\rm {\bf r}}) \big| \psi^{(+)}({\rm {\bf k}},\,{\rm{\bf r}}_{\alpha\,d})\right>\,{\rm {\bf A}}^{*}_{\lambda\,{\rm {\bf k}}_{\gamma}}({\rm {\bf r}})  \nonumber\\
& \approx   \sqrt{ \frac{\hbar}{2\,\pi\,c\,k_{\gamma}}}\,\sum\limits_{L\,M}\,\frac{i^{-L+1}\,k_{\gamma}^{L-1}}{(2\,L+1)!!}\,\sqrt{\frac{(L+1)(2\,L+1)}{L}}\,(D_{M\,\lambda}^{L}(\varphi, \theta,0))^{*}                 \nonumber\\
&\int\,{\rm d}{\rm {\bf r}}\,
\left<I_{l_{f}\,s\,J_{f}}({\rm {\bf r}}_{\alpha\,d})\big|{\rm {\bf {\hat J}}}({\rm {\bf r}}) \big| \psi^{(+)}({\rm {\bf k}},\,{\rm{\bf r}}_{\alpha\,d})\right>                                                                          
{\rm {\bm \nabla}}_{\rm {\bf r}} [r^{L}\,\big(Y_{L\,M}({\rm {\bf {\hat r}}})\big)^{*}] .     
\label{matrelementlongwvlength1}
\end{align}

Integrating by parts and  using the static current conservation 
\begin{align}
{\rm {\bm \nabla}}_{\rm {\bf r}} {\rm {\bf {\hat J}}}({\rm {\bf r}})=  i\,k_{\gamma}\,c\,{\hat \rho}({\rm {\bf r}}),
\label{currconserveqn1}
\end{align}
where 
\begin{align}
{\hat \rho}({\rm {\bf r}})= Z_{d}\,e\,\delta({\rm {\bf r}} - {\rm {\bf r}}_{d}) + Z_{\alpha}\,e\,\delta({\rm {\bf r}}- {\rm {\bf r}}_{\alpha})
\label{elchdensoper1}
\end{align}
is the charge density operator, one gets
\begin{align}
&\frac{1}{c}\int\,{\rm d}{\rm {\bf r}}\,
\left<I_{l_{f}\,s\,J_{f}}({\rm {\bf r}}_{\alpha\,d})\big|{\rm {\bf {\hat J}}}({\rm {\bf r}}) \big|\psi^{(+)}({\rm {\bf k}},\,{\rm{\bf r}}_{\alpha\,d})\right>\,{\rm {\bf A}}^{*}_{\lambda\,{\rm {\bf k}}_{\gamma}}({\rm {\bf r}})  \nonumber\\
&\approx    \frac{1}{2\,\pi}\sqrt{ \frac{\hbar\,c}{2\,k_{\gamma}}}\,\sum\limits_{L\,M}\,\frac{i^{-L}\,k_{\gamma}^{L}}{(2\,L-1)!!}\,\sqrt{\frac{L+1}{L}}\,\big(D_{M\,\lambda}^{L}(\varphi, \theta,0)\big)^{*}                 \nonumber\\
&\times  \left<I_{l_{f}\,s\,J_{f}}({\rm {\bf r}}_{\alpha\,d})\big|\big({\hat Q}_{L\,M}^{(e)}({\rm {\bf r}}_{\alpha\,d})\big)^{*} \big| \psi^{(+)}({\rm {\bf k}},\,{\rm{\bf r}}_{\alpha\,d})\right>.   
\label{matrelementZiegert1}
\end{align}
Here,
\begin{align}
{\hat Q}_{L\,M}^{(e)}({\rm {\bf r}}_{\alpha\,d})= \sqrt{\frac{4\,\pi}{2\,L+1}}\,\int{\rm d}{\rm {\bf r}}\,{\hat \rho}({\rm {\bf r}})\,{r}^{L}\,Y_{LM}({\rm {\bf  {\hat r}}}),
\label{Qe1}
\end{align}
is the electric static $2^{L}$ moment operator.

Thus the initial matrix element (\ref{trmatrelem1})  containing  ${\rm {\bf A}}^{*}_{\lambda\,{\rm {\bf k}}_{\gamma}}({\rm {\bf r}}) $ after the multipole expansion and series of transformations is reduced to the matrix element,  which is expressed in terms of the electric charge density operator.  This is possible due  to  Siegert's theorem \cite{Siegert}. 

Equation (\ref{diffcrsect1})  for the  differential cross section of the electric transition takes the form
\begin{align}
&\frac{{\rm d}\sigma_{\lambda}}{{\rm d}\Omega}  \sim \big| -\frac{1}{2\,\pi}\sqrt{ \frac{\hbar\,c}{2\,k_{\gamma}}}\,\sum\limits_{L\,M}\,\frac{i^{-L}\,k_{\gamma}^{L+1}}{(2\,L-1)!!}\,\sqrt{\frac{L+1}{L}}\,\big(D_{M\,\lambda}^{L}(\varphi, \theta,0)\big)^{*} 
\nonumber\\
&\times\left<I_{l_{f}\,s\,J_{f}}({\rm {\bf r}}_{\alpha\,d})\big|\big({\hat Q}^{(e)}_{L\,M}({\rm {\bf r}}_{\alpha\,d} )\big)^{*}\big| \psi^{(+)}({\rm {\bf k}},\,{\rm {\bf r}}_{\alpha\,d})\right>\Big|^{2}.
\label{diffcrsect2}
\end{align}
In the case under consideration the dominant contribution comes from 
the electric dipole ($L=1$)  and electric quadrupole ($L=2$)  transitions.  Because the sum over multipoles $L$ is incoherent the interference of the dipole and quadrupole amplitudes should be taken into account. 

Integrating over ${\rm {\bf r}}$ in Eq. (\ref{Qe1}) one gets 
\begin{align}
{\hat Q}_{L\,M}^{(e)}({\rm {\bf r}}_{\alpha\,d})= \sqrt{\frac{4\,\pi}{2\,L+1}}\,e\,Z_{eff(L)}\,r_{\alpha\,d}^{L}\,Y_{LM}({\rm {\bf {\hat r}}}_{\alpha\,d}).
\label{Qel2}
\end{align}
$e\,Z_{eff(L)}$ is the effective charge for the electric  transition of the multipolarity $L$, where
\begin{align}
Z_{eff(L)}= \mu_{\alpha\,d}^{L}\left(\frac{Z_{\alpha}}{m_{\alpha}^{L}}  + (-1)^{L}\,\frac{Z_{d} }{m_{d}^{L} }\right).
\label{ZeffL1}
\end{align}
To derive Eq. (\ref{Qel2})  we took into account that  $Y_{L\,M}(-{\rm {\bf {\hat r}}}_{\alpha\,d}) = (-1)^{L}\,Y_{L\,M}({\rm{\bf {\hat r}}}_{\alpha\,d})$.

The improvement of the leading order of the long wavelength approximation leads to the replacement of $r_{\alpha\,d}^{L}$  in Eq. (\ref{Qel2})  by more refined expressions \cite{donnelly}. 
For the dipole transition $r_{\alpha\,d}$ in Eq. (\ref{Qel2})   should be replaced by 
\begin{align}
O_{1}(r_{\alpha\,d}) = \frac{3}{y^{3}}\,\big[(y^{2}- 2)\,\sin y +2 y \cos y \big]\,r_{\alpha\,d}
\label{O11}
\end{align}
and for the quadrupole transition $r_{\alpha\,d}^{2}$ should be replaced by 
\begin{align}
O_{2}(r_{\alpha\,d})= \frac{15}{y^{5}}\,\big[ (5\,y^{2} - 12)\,\sin y + (12 - y^{2})\,y\cos y \big]\,r_{\alpha\,d}^{2},
\label{O21}
\end{align}
where $y=k_{\gamma}\,r_{\alpha\,d}$.

The initial  scattering wave function with  spin-orbit interaction is given by
\begin{align}
&\psi^{(+)}({{\rm {\bf k}},\,\rm {\bf r}}_{\alpha\,d}) = 4\,\pi\,\sum\limits_{J_{i}}\sum\limits_{l_{i}}\,i^{l_{i}}\,\psi_{l_{i}\,sJ_{i}}^{(+)}(k,r_{\alpha\,d})\,
\sum\limits_{m_{l_{i}} m_{s}}\,\left<l_{i}\,m_{l_{i}}\,\,s\,m_{s}|J_{i}\,M_{i}\right>\,Y_{ l_{i}\,m_{l_{i} }}({\rm{\bf {\hat r}}}_{\alpha\,d})       \nonumber\\
& \times \chi_{s\,m_{s}}\, \sum\limits_{m_{l_{i}}' m_{s}' }\,\left<l_{i}\,m_{l_{i}}'\,\,s\,m_{s}'|J_{i}\,M_{i}\right>\,Y_{l_{i}\,m_{l_{i}}' }^{*}({\rm{\bf {\hat k}}}).
\label{psiinit1}
\end{align}
It is assumed that the  projection $M_{i}$ of $J_{i}$  is fixed. For $z\| {\rm {\bf k}}$  $\,Y_{l_{i}\,m_{l_{i}}'}({\rm{\bf {\hat k}}})= \sqrt{{(2\,l_{i}+1)}/{4\,\pi}}\,\delta_{m_{l_{i}}'\,0}$ and, hence, $m_{s}'=M_{i}$. Then
\begin{align}
&\psi^{(+)}({{\rm {\bf k}},\,\rm {\bf r}}_{\alpha\,d}) = \sum\limits_{J_{i}}\sum\limits_{l_{i}}\,i^{l_{i}}\,\sqrt{4\,\pi(2\,l_{i}+1)}\,\psi_{l_{i}\,sJ_{i}}^{(+)}(k,r_{\alpha\,d})\,
\sum\limits_{m_{l_{i}} m_{s}}\,\left<l_{i}\,m_{l_{i}}\,\,s\,m_{s}|J_{i}\,M_{i}\right>\,Y_{ l_{i}\,m_{l_{i} }}({\rm{\bf {\hat r}}}_{\alpha\,d})       \nonumber\\
& \times \chi_{s\,m_{s}}\, \left<l_{i}\,0\,\,s\,M_{i}|J_{i}\,M_{i}\right>.
\label{psiinit2}
\end{align}

The asymptotic behavior of the radial scattering wave function is taken in the form
\begin{align}
\psi_{l_{i}\,sJ_{i}}^{(+)}(k,\,r_{\alpha\,d}) \approx \frac{1}{2\,i\,r_{\alpha\,d}}e^{-i\,\delta_{l_{i} s J_{i}}}\,\big[I_{l_{i}}(k,r_{\alpha\,d})  - e^{2\,i\,\delta_{l_{i}s J_{i}}}\,O_{l_{i}}(k,\,r_{\alpha\,d})\big].
\label{asscatwvf1}
\end{align}
\begin{align}
I_{l_{i}}(k,\,r_{\alpha\,d})=  G_{l_{i}}(k,\,r_{\alpha\,d}) - i\,F_{l_{i}}(k,\,r_{\alpha\,d})
\label{ingspwave1}
\end{align}
and 
\begin{align}
O_{l_{i}}(k,\,r_{\alpha\,d})= G_{l_{i}}(k,\,r_{\alpha\,d})  + i\,F_{l_{i}}(k,\,r_{\alpha\,d})
\label{outgspwave1}
\end{align}
are the incoming and outgoing spherical waves expressed in terms of the regular, $F_{l_{i}}(k,\,r_{\alpha\,d})$, and singular, $G_{l_{i}}(k,\,r_{\alpha\,d})$,
Coulomb solutions of the radial Schr\"odinger equation. $\delta_{l_{i} s J_{i}}$ is the scattering phase shift.  

Inserting Eqs (\ref{overlapfunction1}) and (\ref{psiinit2}) into the matrix element of Eq. (\ref{diffcrsect2}) one finds that
\begin{align}
&\left<I_{l_{f}\,s\,J_{f}}({\rm {\bf r}}_{\alpha\,d})\big|{\hat Q}^{(e)}_{L\,M}({\rm {\bf r}} )\big| \psi^{(+)}({\rm {\bf k}},\,{\rm {\bf r}}_{\alpha\,d})\right> =  \sqrt{\frac{4\,\pi}{2\,L+1}}\,
\sum\limits_{J_{i}}\sum\limits_{l_{i}}\,i^{l_{i}}\,\sqrt{4\,\pi\,(2\,l_{i}+1)}                                                                       \nonumber\\
& \times \sum\limits_{m_{l_{i}} m_{s}\,m_{l_{f}}}\,\left<l_{f}\,m_{l_{f}}\,\,s\,m_{s}\big|J_{f}\,M_{f}\right>\,\left<l_{i}\,m_{l_{i}}\,\,s\,m_{s}|J_{i}\,M_{i}\right>\,\left<l_{i}\,0\,\,s\, M_{i}\big|J_{i}\,M_{i}\right>                                                                           \nonumber\\
& \times \left<I_{l_{f}\,s\,J_{f}}(r_{\alpha\,d})\,Y_{l_{f}m_{l_{f}}}({\rm {\bf {\hat r}}}_{\alpha\,d})\,\big|e\,Z_{eff(L)}\,r_{\alpha\,d}^{L}\,Y_{LM}^{*}({\rm {\bf {\hat r}}}_{\alpha\,d})\big|Y_{ l_{i}\,m_{l_{i} }}({\rm{\bf {\hat r}}}_{\alpha\,d})\,\psi_{l_{i}\,s\,J_{i}}^{(+)}(k,r_{\alpha\,d})\right> \nonumber\\
&= \sqrt{4\,\pi\,(2\,l_{f}+1)}\,e\,Z_{eff(L)}\,\sum\limits_{J_{i}} \sum\limits_{l_{i}}\,\sum\limits_{m_{l_{i}}\,m_{s}\,m_{l_{f}} }i^{l_{i}}\,\left<l_{f}\,m_{l_{f}}\,\,s\,m_{s}\big|J_{f}\,M_{f}\right> \,\left<l_{i}\,m_{l_{i}}\,\,s\,m_{s}|J_{i}\,M_{i}\right>                                                                       \nonumber\\ 
& \times \,\left<l_{i}\,0\,\, s\, M_{i}\big|J_{i}\,M_{i}\right> \, \left<l_{f}0\,L0\big|l_{i}0\right> \,\left<l_{f}\,m_{l_{f}}\,\,L\,M\big|l_{i}\,m_{l_{i}}\right> \,R_{l_{f}\,s\,L\,J_{f}\,l_{i}\,J_{i}}(k),
\label{matrelement2}
\end{align}
\begin{align}
R_{l_{f}\,s\,L\,J_{f}\,l_{i}\,J_{i}}(k) = \,\int_{0}^{\infty}\,{\rm d}r_{\alpha\,d}\,r_{\alpha\,d}^{L+2}\, I_{l_{f}\,s\,J_{f}}(r_{\alpha\,d})\,\psi_{l_{i}\,s\,J_{i}}^{(+)}(k,r_{\alpha\,d}).
\label{matrelemitM1}
\end{align}
	
When deriving Eq. (\ref{matrelement2})  it was taken into account that $\left<\chi_{sm_{s}''}\big|\chi_{sm_{s}}\right> =\delta_{m_{s}''\,m_{s}}$  and \cite{edmonds}
\begin{align}
&\int\,{\rm d}\Omega\,Y_{l_{f}m_{l_{f}}}^{*}({\rm {\bf {\hat r}}}_{\alpha\,d})\,Y_{LM}^{*}({\rm {\bf {\hat r}}}_{\alpha\,d})\,Y_{ l_{i}\,m_{l_{i} }}({\rm{\bf {\hat r}}}_{\alpha\,d})=
 \sqrt{\frac{(2\,l_{f}+1)\,(2\,L+1)}{4\,\pi\,(2\,l_{i} +1)}}\,\left<l_{f}\,0\,\,L\,0\big|l_{i}\,0\right>    \nonumber\\
&\times \left<l_{f}\,m_{l_{f}}\,\,L\,M\big|l_{i}\,m_{l_{i}}\right>.
\label{spherfunctint1}
\end{align}

Now we are able to rewrite the expression for the photon differential cross section including all the kinematical factors. If the polarization of the initial and final nuclei (in the case under consideration deuteron and ${}^{6}{\rm Li}$) and of the photon are not measured  then the differential cross section takes the form
\begin{align}
&\frac{{\rm d}\sigma}{{\rm d}\Omega} = \frac{1}{4}\,\frac{(2\,l_{f}+1)}{(2\,J_{d}+1)(2\,J_{\alpha}+1)}\,\frac{(\hbar\,c)^{3}}{\mu_{\alpha\,d}\,c^{2}}\,\frac{k}{E^{2}}\frac{e^{2}}{\hbar\,c}
\sum\limits_{M_{i}\,M_{f}}\,\sum\limits_{J_{i}'\,J_{i}}\, \sum\limits_{L'L}\,i^{L'-L}\,Z_{eff(L')}\,Z_{eff(L)}\,                                                             \nonumber\\
&\times  \sqrt{\frac{(L'+1)(L+1)}{L'\,L}}\, \frac{k_{\gamma}^{L'+L+1}}{(2\,L'-1)!!\,(2\,L-1)!!} \,\sum\limits_{m_{l_{f}}'\,m_{l_{f}}}\,\sum\limits_{m_{s}'\,m_{s}} \,\sum\limits_{M' \,M}
\sum\limits_{\lambda = \pm 1}\,\,D_{M'\,\lambda}^{L'}(\varphi,\,\theta,0) \,{D_{M\,\lambda}^{{L}^{*}}}(\varphi,\,\theta,0)\,                              \nonumber\\
&\times \sum\limits_{l_{i}'\,l_{i}}\,\sum\limits_{m_{l_{i}}'\,m_{l_{i}}}\,i^{l_{i} - l_{i}'}\,\left<l_{f}\,m_{l_{f}}'\,\,s\,m_{s}'\big|J_{f}\,M_{f}\right> \,\left<l_{f}\,m_{l_{f}}\,\,s\,m_{s}\big|J_{f}\,M_{f}\right>\,\left<l_{i}'\,m_{l_{i}}'\,\,s\,m_{s}'|J_{i}'\,M_{i}\right>                                        \nonumber\\
& \times\left<l_{i}\,m_{l_{i}}\,\,s\,m_{s}|J_{i}\,M_{i}\right>\,\left<l_{i}'\,0\,\, s\, M_{i}\big|J_{i}'\,M_{i}\right>\,\left<l_{i}\,0\,\, s\, M_{i}\big|J_{i}\,M_{i}\right>\,\left<l_{f}\,0\,\,L'\,0\big|l_{i}'0\right>\,\left<l_{f}\,0\,\,L\,0\big|l_{i}\,0\right>                                                                                                              \nonumber\\
& \times \left<l_{f}\,m_{l_{f}}'\,\,L'\,M'\big|l_{i}'\,m_{l_{i}}'\right>\,\left<l_{f}\,m_{l_{f}}\,\, L\,M\big|l_{i}\,m_{l_{i}}\right>\, R_{l_{f}\,s\,L'\,J_{f}\,l_{i}' \,J_{i}} ^{*}(k)\,R_{l_{f}\,s\,L\,J_{f}\,l_{i}\,J_{i}}(k).
\label{diffcrsect3}
\end{align}

Equation (\ref{diffcrsect3})  can be further simplified taking into account that \cite{Varshalovich}
\begin{align}
&\sum\limits_{m_{l_{f}}\,m_{s}\,m_{l_{i}}} \left<l_{f}\,m_{l_{f}}\,\,s\,m_{s}\big|J_{f}\,M_{f}\right>\,\left<l_{i}\,m_{l_{i}}\,\,s\,m_{s}|J_{i}\,M_{i}\right>\,\left<l_{f}\,m_{l_{f}}\,\, L\,M\big|l_{i}\,m_{l_{i}}\right>  \nonumber\\
&= (-1)^{s+J_{f}+l_{i}+ L}\,\sqrt{(2\,J_{f}+1)(2\,l_{i}+1)}\,\left<J_{f}\,M_{f}\,\,L\,M \big| J_{i}\,M_{i}\right>\,
\Bigg\lbrace \begin{array}{ccc} l_{f} \,s\,J_{f} \\
 J_{i}\,L\,l_{i}  \end{array} \Bigg\rbrace,
\label{Rakachcoef1}
\end{align}
where $\Bigg\lbrace \begin{array}{ccc} l_{f} \,s\,J_{f} \\
 J_{i}\,L\,l_{i}  \end{array} \Bigg\rbrace$  is the $6j$-symbol \cite{Varshalovich}.

Then
\begin{align}
&\frac{{\rm d}\sigma}{{\rm d}\Omega} = \frac{1}{4}\,\frac{(2\,l_{f}+1)(2\,J_{f}+1)}{(2\,J_{d}+1)(2\,J_{\alpha}+1)}\,\frac{(\hbar\,c)^{3}}{\mu_{\alpha\,d}\,c^{2}}\,\frac{k}{E^{2}}\frac{e^{2}}{\hbar\,c}
\sum\limits_{M_{i}\,M_{f}}\,\sum\limits_{J_{i}'\,J_{i}}\, \sum\limits_{L'L}\,i^{L'-L}\,Z_{eff(L')}\,Z_{eff(L)}\,                                                             \nonumber\\
&\times  \sqrt{\frac{(L'+1)(L+1)}{L'\,L}}\, \frac{k_{\gamma}^{L'+L+1}}{(2\,L'-1)!!\,(2\,L-1)!!} \,\sum\limits_{M' \,M}
\sum\limits_{\lambda = \pm 1}\,\,D_{M'\,\lambda}^{L'}(\varphi,\,\theta,0) \,{D_{M\,\lambda}^{{L}^{*}}}(\varphi,\,\theta,0)\,                              \nonumber\\
&\times \sum\limits_{l_{i}'\,l_{i}}\,i^{l_{i} - l_{i}'}\,\left<l_{i}'\,0\,\, s\, M_{i}\big|J_{i}'\,M_{i}\right>\,\left<l_{i}\,0\,\, s\, M_{i}\big|J_{i}\,M_{i}\right>\,\left<l_{f}\,0\,\,L'\,0\big|l_{i}'0\right>\,\left<l_{f}\,0\,\,L\,0\big|l_{i}\,0\right>    \nonumber\\
& \times (-1)^{ l_{i}'+l_{i}+ L'+L}\,\sqrt{(2\,l_{i}'+1)(2\,l_{i}+1)}\,\left<J_{f}\,M_{f}\,\,L'\,M' \big|J_{i}'\,M_{i}\right>\,\left<J_{f}\,M_{f}\,\, L\,M \big| J_{i}\,M_{i}\right>    \nonumber\\& \times  \Bigg\lbrace \begin{array}{ccc} l_{f} \,s\,J_{f} \\
 J_{i}'\,L'\,l_{i}'  \end{array} \Bigg\rbrace\,\,\Bigg\lbrace \begin{array}{ccc} l_{f} \,s\,J_{f} \\
 J_{i}\,L\,l_{i}  \end{array} \Bigg\rbrace\,R_{l_{f}\,s\,L'\,J_{f}\,l_{i}' \,J_{i}'} ^{*}(k)\,R_{l_{f}\,s\,L\,J_{f}\,l_{i}\,J_{i}}(k).
\label{diffcrsect4}
\end{align}
From $\left<J_{f}\,M_{f}\,\,L'\,M' \big|J_{i}'\,M_{i}\right>\,\left<J_{f}\,M_{f}\,\, L\,M \big| J_{i}\,M_{i}\right> $  follows that $M'=M$ and \cite{edmonds}
\begin{align} 
&D_{M\,\lambda}^{{L' }} (\varphi,\,\theta,0)\,\,\big(D_{M\,\lambda}^{L}(\varphi,\,\theta,0)\big)^{*} = (-1)^{M- \lambda}\,
D_{M\,\lambda}^{L'} (\varphi,\,\theta,0)\,D_{-M\,-\lambda}^{L}(\varphi,\,\theta,0)                                                                         \nonumber\\
&= (-1)^{M-\lambda}\,\sum\limits_{J}\,\left<L' M\,\,L -M\big|J0\right>\,\left<L' \lambda\,\,L -\lambda \big|J0\right>\,
D_{00}^{J}(\varphi,\,\theta,0)                                                                                                                                               \nonumber\\
&= -(-1)^{M}\,\sum\limits_{J}\,\left<L' M\,\,L -M\big|J0\right>\,\left<L' \lambda\,\,L -\lambda \big|J0\right>\,P_{J}(\cos\theta)
\label{Dfunctions1}
\end{align}
and
\begin{align}
&\sum\limits_{\lambda= \pm 1}\,\left<L' \lambda\,\,L -\lambda \big|J0\right>=  \left<L' 1\,\,L -1 \big|J0\right>\big[1+ (-1)^{L'+L-J}\big].
\label{ClebGordsymmetry1}
\end{align}

Then
\begin{align}
&\frac{{\rm d}\sigma}{{\rm d}\Omega} = -\frac{1}{4}\,\frac{(2\,l_{f}+1)(2\,J_{f}+1)}{(2\,J_{d}+1)(2\,J_{\alpha}+1)}\,\frac{(\hbar\,c)^{3}}{\mu_{\alpha\,d}\,c^{2}}\,\frac{k}{E^{2}}\frac{e^{2}}{\hbar\,c}
\sum\limits_{M_{i}\,M_{f}}\,\sum\limits_{J_{i}'\,J_{i}}\, \sum\limits_{L'L}\,i^{L'-L}\,Z_{eff(L')}\,Z_{eff(L)}\, \sqrt{\frac{(L'+1)(L+1)}{L'\,L}}\,                                                             \nonumber\\
&\times  \frac{k_{\gamma}^{L'+L+1}}{(2\,L'-1)!!\,(2\,L-1)!!} \,\sum\limits_{M}\,(-1)^{M}\,\sum\limits_{J}\,
\left<L' M\,\,L -M\big|J\,0\right>\,\left<L' 1\,\,L -1 \big|J\,0\right>\big[1+ (-1)^{L'+L-J}\big]\                                     \nonumber\\
&\times P_{J}(\cos\theta) \,\sum\limits_{l_{i}'\,l_{i}}\,i^{l_{i} - l_{i}'}\,\left<l_{i}'\,0\,\, s\, M_{i}\big|J_{i}'\,M_{i}\right>\,\left<l_{i}\,0\,\, s\, M_{i}\big|J_{i}\,M_{i}\right>\,\left<l_{f}\,0\,\,L'\,0\big|l_{i}'0\right>\,\left<l_{f}\,0\,\,L\,0\big|l_{i}\,0\right>    \nonumber\\
& \times (-1)^{ l_{i}'+l_{i}+ L'+L}\,\sqrt{(2\,l_{i}'+1)(2\,l_{i}+1)}\,\left<J_{f}\,M_{f}\,\,L'\,M \big|J_{i}'\,M_{i}\right>\,\left<J_{f}\,M_{f}\,\, L\,M \big| J_{i}\,M_{i}\right>    \nonumber\\& \times  \Bigg\lbrace \begin{array}{ccc} l_{f} \,s\,J_{f} \\
 J_{i}'\,L'\,l_{i}'  \end{array} \Bigg\rbrace\,\,\Bigg\lbrace \begin{array}{ccc} l_{f} \,s\,J_{f} \\
 J_{i}\,L\,l_{i}  \end{array} \Bigg\rbrace\,\, R_{l_{f}\,s\,L'\,J_{f}\,l_{i}' \,J_{i}'} ^{*}(k)\,R_{l_{f}\,s\,L\,J_{f}\,l_{i}\,J_{i}}(k).
\label{dcrsctgeneral1}
\end{align}
Eq. (\ref{dcrsctgeneral1}) is quite general and can be applied for the analysis of the photon angular distribution in the direct radiative capture reactions contributed  by electric transitions with different multipolarities $L$ or with one dominant $L$.  
In Eq. (\ref{diffcrsect3})  $\,\hbar\,c= 197.3$ MeV\,fm,  $\,e^{2}/(\hbar\,c)=1/137$,  $\mu_{\alpha\,d}\,c^{2}$ and $E$  are expressed in MeV,  $k_{\gamma}$ and $k$ are expressed in fm${}^{-1}$.  
Assuming that only $L=1$ or $L=2$ contribute one can easily derive differential cross sections for the electric dipole and quadrupole transitions.

Equation (\ref{dcrsctgeneral1})  can be further simplified for the ${}^{2}{\rm H}(\alpha,\,\gamma){}^{6}{\rm Li}$  reaction, for which $l_{f}=0,\,J_{f}=1,\,s=1,\,J_{\alpha}=0$.  
For this reaction
\begin{align}
&\Bigg\lbrace \begin{array}{ccc} 0 \,s\,J_{f} \\
 J_{i}\,L\,l_{i}  \end{array} \Bigg\rbrace  =  (-1)^{J_{f}+ L+J_{i}}\, \frac{\delta_{s\,J_{f}}\,\delta_{l_{i}\,L} }{\sqrt{(2\,J_{f}+1)(2\,L+1)}},    \nonumber\\
&\Bigg\lbrace \begin{array}{ccc} 0 \,s\,J_{f} \\
 J_{i}'\,L'\,l_{i}'  \end{array} \Bigg\rbrace   =  (-1)^{J_{f}+ L'+J_{i}'}\, \frac{\delta_{s\,J_{f}}\delta_{l_{i}'\,L'} }{\sqrt{(2\,J_{f}+1)(2\,L'+1)}},
\label{simplify6jymb1}
\end{align}
$\left<0\,0\,\,L\,0\big|l_{i}0\right> =  \delta_{l_{i}\,L}$ and $\left<0\,0\,\,L'\,0\big|l_{i}'0\right> =  \delta_{l_{i}'\,L'}$.   

Then for the differential cross section for the reaction under consideration we get
\begin{align}
&\frac{{\rm d}\sigma}{{\rm d}\Omega} =- \frac{1}{12}\,\frac{(\hbar\,c)^{3}}{\mu_{\alpha\,d}\,c^{2}}\,\frac{k}{E^{2}}\frac{e^{2}}{\hbar\,c}
\sum\limits_{M_{i}\,M_{f}}\,\sum\limits_{J_{i}'\,J_{i}}\, \sum\limits_{L'L}\,Z_{eff(L')}\,Z_{eff(L)}\, \sqrt{\frac{(L'+1)(L+1)}{L'\,L}}\,                                                             \nonumber\\
&\times \frac{k_{\gamma}^{L'+L+1}}{(2\,L'-1)!!\,(2\,L-1)!!} \,\sum\limits_{M}\,(-1)^{M}\,\sum\limits_{J}\,
\left<L' M\,\,L -M\big|J0\right>\,\left<L' 1\,\,L -1 \big|J0\right>                                 \nonumber\\
&\times   \big[1+ (-1)^{L'+L-J}\big]\,P_{J}(\cos\theta) \,\left<L'\,0\,\,J_{f}\, M_{i}\big|J_{i}'\,M_{i}\right>\,\left<L\,0\,\, J_{f}\, M_{i}\big|J_{i}\,M_{i}\right>    \nonumber\\
& \times \left<J_{f}\,M_{f}\,\,L'\,M \big|J_{i}'\,M_{i}\right>\,\left<J_{f}\,M_{f}\,\, L\,M \big| J_{i}\,M_{i}\right> \,R_{0\,L'\,1\,J_{i}'} ^{*}(k)\,R_{0\,L\,1\,J_{i}}(k).
\label{diffcrsectgeneral6Li1}
\end{align}

Equations (\ref{dcrsctgeneral1}) and (\ref{diffcrsectgeneral6Li1}) are our first main result.

\subsection{Total cross sections}
\label{total cross sections 1}

The total cross sections can be obtained by integrating the above differential cross sections over the photon's solid angle.  
Integrating Eq. (\ref{dcrsctgeneral1})  keeps only the term $J=0$  what leads to $L'=L$. 
Then
\begin{align}
\sum\limits_{M_{f}\,M} \left<J_{f}\,M_{f}\,\,L\,M \big|J_{i}'\,M_{i}\right>\,\left<J_{f}\,M_{f}\,\, L\,M \big| J_{i}\,M_{i}\right>= \delta_{J_{i}\,J_{i}},
\label{ClebschGordanorth1}
\end{align}
$\left<L M\,\,L -M\big|0\,0\right>=(-1)^{L-M}\,\sqrt{{1}/({2\,L+1})}$ and $\left<L 1\,\,L -1 \big|0\,0\right>= (-1)^{L-1}\,\sqrt{{1}/({2\,L+1})}$.
From $\left<l_{f}\,0\,\,L\,0\big|l_{i}\,0\right>$  follows that two subsequent $l_{i}$ can differ by 2. At astrophysically relevant energies 
only minimal $l_{i}$ dominate. Hence we can drop the sum over $l_{i}$ assuming that each $l_{i}$ is uniquely determined by $L$.
Also 
\begin{align}
\left<l_{i}\,0\,\,s\,M_{i} \big| J_{i}\,M_{i}\right>= (-1)^{s+ M_{i}}\,\sqrt{\frac{2\,J_{i}+1}{2\,l_{i}+1}}\,\left<J_{i}\,-M_{i}\,\,s\,M_{i}\big|l_{i}\,0\right>
\label{CLGord2}
\end{align}
and 
\begin{align}
\sum\limits_{M_{i}} \left(\left<J_{i}\,-M_{i}\,\,s\,M_{i}\big|l_{i}\,0\right>\right)^{2}=1.
\label{ClGord3}
\end{align}

Taking into account the above results the total cross section reduces to
\begin{align}
& \sigma = 2\,\pi\,\frac{(2\,l_{f}+1)(2\,J_{f}+1)}{(2\,J_{d}+1)(2\,J_{\alpha}+1)}\,\frac{(\hbar\,c)^{3}}{\mu_{\alpha\,d}\,c^{2}}\,\frac{k}{E^{2}}\frac{e^{2}}{\hbar\,c}\, \sum\limits_{J_{i}} (2\,J_{i}+1)\, \sum\limits_{L}\,\big(Z_{eff(L)}\big)^{2}\, \frac{(L+1)(2\,L+1)}{L}                                                        \nonumber\\
&\times  \frac{k_{\gamma}^{2\,L+1}}{\big((2\,L+1)!!\big)^{2}}\,(\left<l_{f}\,0\,\,L\,0\big|l_{i}0\right> )^{2}\, \Bigg[\Bigg\lbrace \begin{array}{ccc} l_{f} \,s\,J_{f} \\
 J_{i}\,L\,l_{i}  \end{array} \Bigg\rbrace\Bigg]^{2}\,  \big|R_{l_{f}\,s\,L\,J_{f}\,l_{i} \,J_{i}} (k) \big|^{2}.
\label{totcrsectgeneral1}
\end{align}
The total cross section for the dipole (quadrupole) transition can be  obtained from Eq. (\ref{totcrsectgeneral1}) by taking $L=1$ ($L=2$).

The total cross section for the reaction under consideration takes the form ($l_{f}=0,\,s=J_{f},\,l_{i}=L$)
\begin{align}
& \sigma = \frac{2\,\pi}{3}\,\frac{(\hbar\,c)^{3}}{\mu_{\alpha\,d}\,c^{2}}\,\frac{k}{E^{2}}\frac{e^{2}}{\hbar\,c}\, \sum\limits_{J_{i}} (2\,J_{i}+1)\, \sum\limits_{L}\,\big(Z_{eff(L)}\big)^{2}\, \frac{L+1}{L}\,\frac{k_{\gamma}^{2\,L+1}}{\big((2\,L+1)!!\big)^{2}}\, \big|R_{0\,L\,1\,J_{i}} (k) \big|^{2}.
\label{totcrsect6Li1}
\end{align}
Equations  (\ref{totcrsectgeneral1}) and (\ref{totcrsect6Li1}) are our second main result.

The astrophysical factor is determined by
\begin{align}
S(E) = E\,e^{2\,\pi\,\eta_{i}}\,\sigma(E).
\label{astrfactor1}
\end{align}
Here, $\eta_{i}$ is the Coulomb parameter in the initial state of the radiative capture process.  Replacing $\sigma(E)$ by $\sigma_{Ei}(E)$, where $i=1,2$, we get the astrophysical factors for the dipole ($E1$) and quadrupole ($E2$) transitions, correspondingly.

\subsection{Potential model}
\label{potential model1}

The most important quantity in calculations of the radiative capture reactions is the radial matrix element  $R_{l_{f}\,s\,L\,\,J_{f}\,l_{i}\,J_{i}}(k)$, which is expressed in terms of the the initial and final nuclear wave functions.  Different approaches were used to calculate the radial matrix elements.  The most frequent used potential approach was based on the pioneering works  \cite{christyduck,tombrello}. In the potential approach the initial scattering wave function is a solution of the Schr\"odinger equation with the $\alpha-d$ potential, which can be found from the fitting experimental elastic scattering phase shifts in the corresponding partial waves ($l_{i}=1,2$ in the case under consideration).  The result is very sensitive to the choice of the final overlap function  $I_{l_{f}\,s\,J_{f}}(r_{\alpha\,d})$.  It was long ago recognized \cite{muk95} that the ${{}^{2}{\rm H}(\alpha,\gamma){}^{6}{\rm Li}}$ reaction is peripheral at astrophysically relevant energies, that is, 
the overall normalization of the astrophysical factor  at  Big Bang energies $30 \lesssim E \lesssim 400$ keV  is practically determined by the square of the ANC $C_{l_{f}sJ_{f}}$.  

In \cite{ryzhikh}  the ${}^{6}{\rm Li}$  bound-state wave function was calculated within the framework of the multi-cluster dynamic model. Projection of this bound-state wave function on the two-body channel $\alpha + d$ channel gives the overlap function with correct tail.  The two-body potential model was used in \cite{muk95}  to calculate the astrophysical factors for the electric dipole and quadrupole transitions and the total $S(E)$ factor at energies $E  \leq 500$ keV.  In the two-body  potential model the overlap function is replaced by the $\alpha-d$ bound-state wave function:
\begin{align}
I_{l_{f}sJ_{f}}(r_{\alpha\,d}) = S_{n_{r}l_{f}sJ_{f}}^{1/2}\,\varphi_{n_{r} l_{f}sJ_{f}}(r_{\alpha\,d}), 
\label{singlparwf1}
\end{align}
where $\varphi_{n_{r}l_{f}sJ_{f}}(r_{\alpha\,d})$ is the $\alpha-{\rm d}$ two-body bound-state wave function calculated in some phenomenological Woods-Saxon $\alpha-{\rm d}$ plus Coulomb potential, $n_{r}=1$ is the principal quantum number showing the number of the nodes of the radial bound-state wave function at $r_{\alpha\,d} >0$. $\,S_{n_{r}l_{f}sJ_{f}}$ is the spectroscopic factor of the configuration $\alpha +{\rm d}$ in the ground state of ${}^{6}{\rm Li}$. The  tail of the bound-state wave function is given by
\begin{align}
\varphi_{n_{r}l_{f}sJ_{f}}(r_{\alpha\,d}) \stackrel{r_{\alpha\,d}> r_{0}}{\approx} b_{n_{r}l_{f}sJ_{f}}\,W_{-\eta,\,l_{f}+1/2}(2\,\kappa\,r_{\alpha\,d})/r_{\alpha\,d},    
\label{radbstwfas1}
\end{align}
where  $b_{n_{r}l_{f}sJ_{f}}$ is the single-particle ANC. The value of  $b_{n_{r}l_{f}sJ_{f}}$  depends on the adopted bound-state potential.  The spectroscopic factor $S_{n_{r}l_{f}sJ_{f}}$ reflects the fact that the overlap function is not an eigenfunction of any Hamiltonian and, hence, is not normalized to unity, in contrast to the bound-state wave function. Eq. (\ref{radbstwfas1}) puts limitation on the spectroscopic factor for given $b_{n_{r}l_{f}sJ_{f}}$.  

The bound-state  Woods-Saxon potential should be adjusted to obtain the experimental $\alpha-{\rm d}$ binding energy (well-depth procedure). However, there are infinite number of such potentials because there are three  fitting parameters: geometrical parameters, radius and diffuseness, and the well depth.  The final adjustment can be done using the spectroscopic factor. The two-body potential model was also used in \cite{hammache}. To find the $\alpha-{\rm d}$ bound-state wave function the Woods-Saxon potential   was adjusted to fit the experimental $s$-wave  elastic scattering phase shift and to reproduce the experimental $\alpha-{\rm d}$ binding energy.  Since the experimental elastic scattering phase shift includes the many-body effects
of the scattered nuclei, the same is true for the two-body potential, which fits the elastic scattering data.  Hence,  the spectroscopic factor  in Eq. (\ref{radbstwfas1})  should be set to $S_{n_{r}l_{f}sJ_{f}}=1$.  However, there  is  again infinite number of the Woods-Saxon potentials, which differ by the most crucial quantity - the ANC (the inverse scattering problem theorem by Gel'fand-Levitan-Marchenko  \cite{chadansabatier}).  The potential adopted in \cite{hammache} was one of the infinite set of the phase-equivalent potentials  with the  ANC, which exceeds the experimental ANC \cite{blokh93} and {\it ab initio} calculations \cite{navratil}  by $ \approx 18\%$ .  Hence,  the normalization of the peripheral part of the $S(E)$ factor calculated  in \cite{hammache}  exceeded the correct one by $\approx 38\%$.  All these questions about ambiguity of the two-body bound-state potentials were addressed in details in \cite{muk2011}.  

The first full microscopic 6-body approach to calculate the final state ${}^{6}{\rm Li}$ bound-state wave function was developed in \cite{nollett2001}
using the variational Monte Carlo method.  The projection of the ${}^{6}{\rm Li}$ on the two-body channel $\alpha+{\rm d}$ 
has correct  tail with the ANC close to the experimental one \cite{blokh93}.  The calculated  total $S(E)$ factor  is  in a good agreement with direct measurements around  $3^{+}$ resonance at $E=712$ keV.   

In hour work, to calculate the photon differential cross sections we used the potential model approach. To calculate the bound-state wave function, two different potentials were used. The first one is  the Woods-Saxon potential with the geometrical parameters: radius $r_{0}=1.20$ fm and diffuseness $a=0.7$ fm. The square of the  single-particle ANC of the bound-state wave function generated by this potential is  $b_{1011}^{2}=7.22$ fm${}^{-1}$. To get the correct normalization of the leading asymptotic term of the final-state overlap function $I_{011}(r_{\alpha\,d}) $, that is, the square of the ANC  $ C_{011}^{2}=5.29$ fm${}^{-1}$, we have to introduce in Eq. (\ref{singlparwf1})  the  spectroscopic factor  $S_{1011}=0.72$.  This method is referred to as $M1$. 
The second method is similar to the one  described in \cite{muk2011}. In this method, referred to as $M2$, the Woods-Saxon potential used in \cite{hammache}  was modified to generate the bound-state wave function with correct asymptotic behavior. In this case the spectroscopic factor  is $S_{1011}=1$, that is, the overlap function $I_{011}(r_{\alpha\,d}) $  and bound-state wave function  $\,\varphi_{1011}(r_{\alpha\,d})$ do coincide at all radii.
Thus, both used overlap functions have the same asymptotic behavior being different  in the internal region.
In both methods the initial $\,\alpha-{\rm d}$  scattering wave function is generated by the Woods-Saxon potential from \cite{hammache}. 
Its parameters are adjusted to reproduce the experimental phase shifts in the partial waves $l_{i}=1,2$: the radial parameter is $r=1.25$ fm, diffuseness $a=0.65$ fm, the depth of the potential $56.7$ MeV.  At $l_{i}=2$  this potential reproduces the $3^{+}$ resonance. To calculate the bound-state and scattering wave functions and the radial matrix elements we used the modified RADCAP code \cite{RADCAP}.

\section{Photon angular distribution in direct radiative capture ${\bm{{}^{2}{\rm H}(\alpha,\,\gamma){}^{6}{\rm Li}}}$}

\begin{figure}
\includegraphics[scale=0.7,trim={0 0 0 12cm},clip]{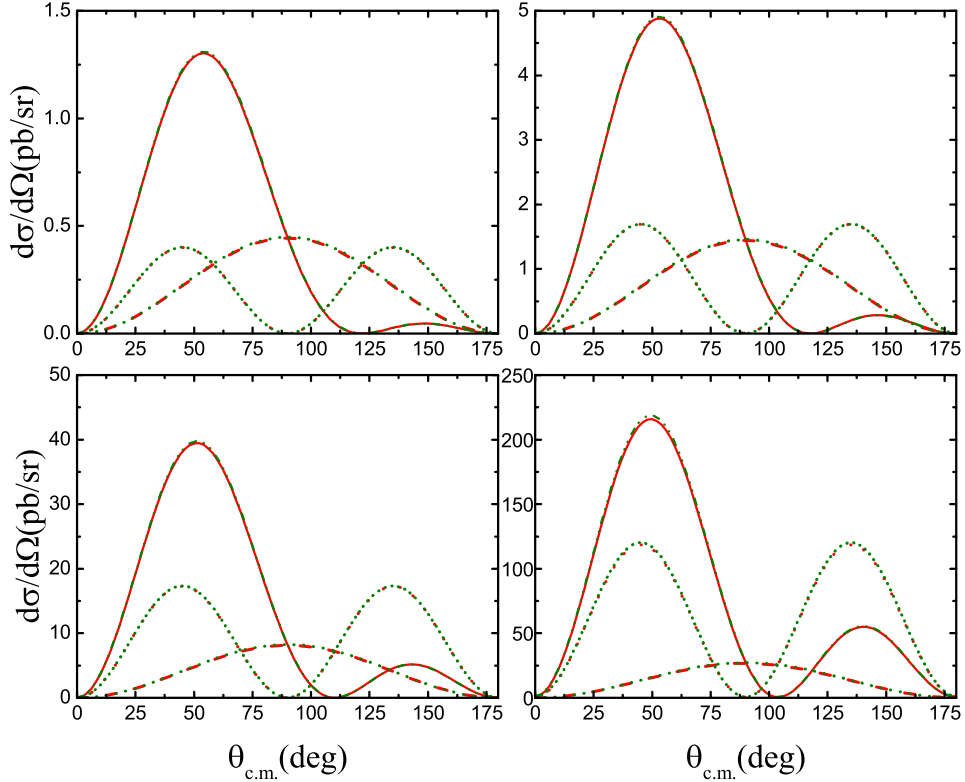}
\caption{(Color online) Angular distributions of the photons emitted in the direct radiative capture  ${}^{2}{\rm H }(\alpha,\gamma){}^{6}{\rm Li}$ at $E=70$ keV (panel (a)),  $E=100$ keV (panel (b)), $E=200$ keV (panel (c)) and $E=400$ keV (panel (d)).  All red (olive) lines are obtained using method $M1$ ($M2$).
The red dashed line (short dashed olive line):  the angular distribution calculated for the $E1$ transition;  the red dotted line (short-dotted olive line): the $E2$ transition; the solid red line (dashed-dotted-dotted olive line): the total photon differential cross section, which is contributed by the  sum of the electric dipole and quadrupole terms and their interference term. } 
\label{fig_angdistr}
\end{figure}

The calculated photon angular distributions for the ${}^{2}{H}(\alpha,\gamma){}^{6}{\rm Li}$ direct radiative capture using both methods, $M1$ and $M2$,  are shown in Fig \ref{fig_angdistr} for 4 different Big Bang energies, $E= 70,\,100,\,200$ and $400$ keV.   
As one can see, the dipole differential cross section has the peak at $90^{\circ}$.  The quadrupole transition has two peaks, at $45^{\circ}$ and $135^{\circ}$.
However their interference dramatically changes  the angular distribution  generating one peak at $\approx 50^{\circ}$. Note that the exact location of the peak slightly depends on the energy. These calculations provide a recipe for the best
experimental kinematics. Note that in the experiment  performed by LUNA \cite{LUNA}  the  germanium detector was placed at a $90^{\circ}$ angle with respect to the ion beam direction. At this angle the differential cross section is significantly smaller than at the peak value at $\approx 50^{\circ}$.

Another important conclusion is that both methods, $M1$ and $M2$, give practically indistinguishable results confirming that at low energies  the reaction  ${}^{2}{H}(\alpha,\gamma){}^{6}{\rm Li}$ is completely peripheral. It means that only the tail of the $\alpha-{\rm d}$  bound-state wave function  contributes to the reaction matrix element.
Hence, to calculate the reaction matrix element it is enough to use any reasonable bound-state Woods-Saxon potential, which supports $s$-wave $\alpha-{\rm d}$ bound state with the $1.47$ MeV binding energy,  and then to introduce a proper spectroscopic factor to provide correct normalization of the asymptotic term of the overlap function.

\section{Astrophysical factor}

\begin{figure}
\includegraphics[scale=0.7,trim={0 0 0 12cm},clip]{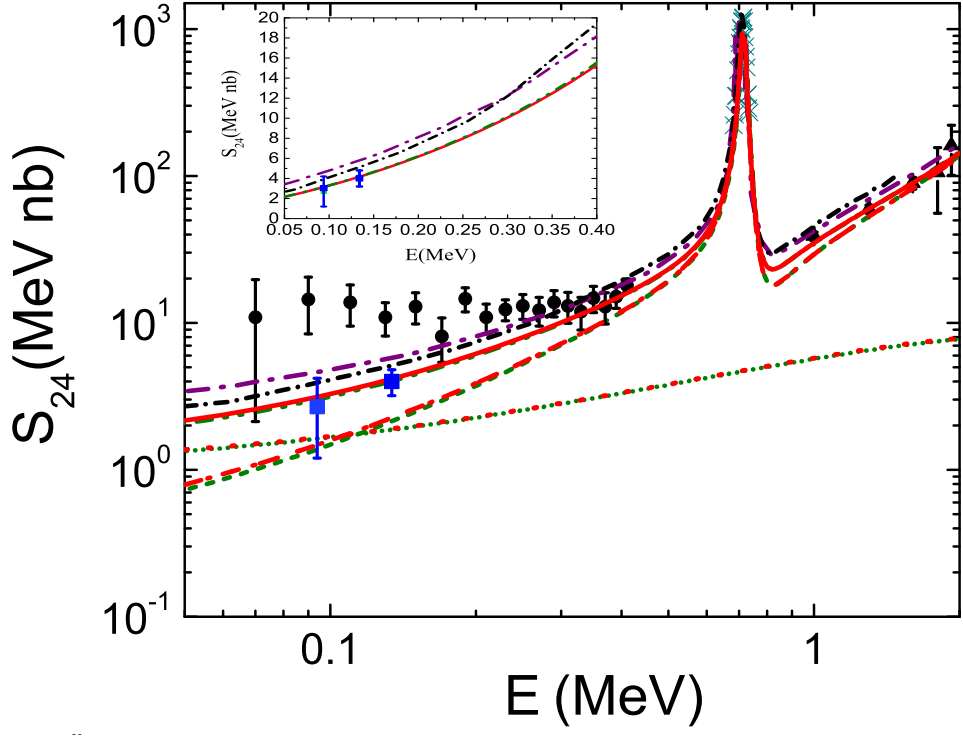}
\caption{(Color online) Astrophysical $S_{24}(E)$ factors for the  ${}^{2}{\rm H}(\alpha,\gamma){}^{6}{\rm Li}$  reaction. 
Black dots are data from  Ref.  \cite{kiener}; black crosses are data from Ref. \cite{mohr};  black triangles are data from Ref. \cite{robertson}.
Two blue boxes are the LUNA experimental  data reported at $E=94$ and $134$ keV \cite{LUNA}  shown together with their  uncertainties.
The purple dashed-dotted line is the $S_{24}(E)$  astrophysical factor from Ref. \cite{nollett}.  The black dashed-dotted line is the $S_{24}(E)$  factor from Ref. \cite{hammache}.  All the red (olive) lines are our calculations obtained using model $M1$ ($M2$). 
The  red dotted (olive short dotted), red dashed (olive short dashed) and red solid (olive dashed-dotted-dotted) lines are the  dipole, quadrupole and total $S_{24}(E)$ factors, correspondingly, from the present  calculations. Notations in the insert are the same.} 
\label{fig_sfactor}
\end{figure}

In Fig. \ref{fig_sfactor}  the  experimental and calculated astrophysical $S_{24}(E)$ factors for the reaction ${}^{2}{\rm H}(\alpha,{\rm d}){}^{6}{\rm Li}$ are presented.
In contrast to the differential cross section, the total astrophysical factor is given by the sum of the dipole and quadrupole astrophysical factors and does not contain their interference term.  
The potential model used in the present calculations with two different bound-state wave functions has been described in section \ref{potential model1}.  The expression for the astrophysical factor  has been derived in section \ref{total cross sections 1}   by integrating the photon's  differential cross section over the photon's solid angle. 
Agreement between the LUNA data at two Big Bang energies and the potential model calculations based on the ANC  provides a compelling evidence of the power of the ANC method. Note that the LUNA results are the first direct measurement of the ${}^{2}{\rm H}(\alpha,\gamma){}^{6}{\rm Li}$  cross section inside the Big Bang  energy range. 

Potential model, which we use here,  allows us to reproduce the available direct data in the region of the first resonance,
$E=0.712$ MeV,  and even at higher energies.  The validity of the potential model at  higher energies can be easily explained. At energies higher than Big Bang energies  the quadrupole transition dominates. Owing to the presence of the factor $r_{\alpha\,d}^{2}\,$  the quadrupole radial matrix element  is dominantly peripheral in the energy interval up to  $\sim 2$ MeV.  Hence,  the used here potential model with  correct normalization of the tail of the overlap function $I_{011}(r_{\alpha\,d}) $ given by the ANC allows one to calculate the astrophysical factor in the broad energy interval $0 \leq E \leq 2$ MeV.   

Note that the calculations from \cite{nollett} and \cite{hammache} at Big Bang energies are higher then presented here.  For example, at $70$ keV, which is the most effective Big Bang energy,  $\,S_{24}(70 {\rm keV}) = 4.0$  MeV nb \cite{nollett},  $S_{24}(70 {\rm keV}) = 3.16$ MeV nb  \cite{hammache}  and  the present result is $S_{24}(70 {\rm keV})=  2.58$ MeV nb.  The insert in Fig. \ref{fig_sfactor} shows the difference between  different calculations of the  $S_{24}(E)$ factors in the Big Bang energy interval. At higher energies calculations from \cite{nollett} reproduce the data quite well while the results  from \cite{hammache} are systematically higher than the data before and after the resonance. 

The accuracy of the long wavelength approximation in the case under consideration is quite high:  a replacement of $r^{L}$ in the integrand of the  radial matrix elements (\ref{matrelemitM1}) by  $O_{1}(r)$,  Eq. (\ref{O11}),
for $L=1$ and $O_{2}(r)$, Eq. (\ref{O21}), for $L=2$ changes the astrophysical factor by only  $\approx 1\%$. 
Note that two data points obtained by LUNA were extrapolated in \cite{LUNA} to other energies using calculations in \cite{muk2011}. The calculations in this paper using 
the method $M2$ are similar to calculations from \cite{muk2011} but performed with a different, more accurate code \cite{RADCAP}. 

Hence  the reaction rates 
calculated here  and in \cite{LUNA}  also agree. These reaction rates are significantly lower than the  adopted reaction rate  from \cite{caughlanfowler} and systematically lower than the reaction rate adopted by NACRE \cite{angulo}. For example, at  $T_{9}=1$, which corresponds to $E= 86.2$ keV,  the adopted NACRE reaction rate exceeds  the calculated one in \cite{muk2011} by about $21.5\%$.

\section{${}^{6}{\rm Li}/{}^{7}{\rm Li}$ isotopic primordial abundance ratio}

Evidently that the  present paper and LUNA's estimations of the Big Bang abundance of ${}^{6}{\rm Li}$ based on the reaction rate of the ${}^{2}{\rm H}(\alpha,\gamma){}^{6}{\rm Li}$  coincide. For the baryon-to-photon ratio  $6.047\times 10^{-10}$, which is within the interval determined by the Planck collaboration \cite{Planck},  the calculated primordial  abundance of  ${}^{6}{\rm Li}$   is 
${}^{6}{\rm Li/H} = (0.74 \pm 0.16)\times 10^{-14}$ \cite{LUNA} which is $34\%$ lower than the abundance given in \cite{caughlanfowler}.

In the latest comprehensive analysis of the Big Bang nucleosynthesis   the primordial abundance of ${}^{6}{\rm Li}$ was determined to be ${}^{6}{\rm Li}/{\rm H}=( 0.90 - 1.77)\times 10^{-14}$ ( Planck baryon-to-photon ratio was adopted)  \cite{coc2014}  and 
${}^{6}{\rm Li}/{\rm H}=(1.23- 1.32) \times 10^{-14}$ (WMAP baryon-to-photon ratio was taken into account) \cite{Coc2012}. As we see, the central values of both  results  are twice as high as LUNA and present estimations. In both works \cite{coc2014,Coc2012}  the nuclear reaction rate from \cite{hammache}  was used  claiming that this reaction rate was obtained from the ${}^{6}{\rm Li}$ Coulomb breakup. However, it was clearly stated in \cite{hammache} that the attempt to determine the $S_{24}(E)$ factor from the Coulomb breakup failed and that a potential two-body model was used to calculate $S_{24}(E)$, which turns out to be  $\sim 30\%$ higher
 than our and LUNA astrophysical factors \cite{muk2011} because a too large value of the ANC was used in \cite{hammache}. Hence, the second claim in \cite{coc2014}  that  the calculated astrophysical factor in \cite{hammache} and experimental  LUNA astrophysical factor  \cite{LUNA} ``agree well" is also questionable  and one of the reasons of high values of the ${}^{6}{\rm Li}$ primordial abundance  obtained  in \cite{coc2014,Coc2012} is that  the adopted reaction rates  for the ${}^{2}{\rm H}(\alpha,\gamma){}^{6}{\rm Li}$ were based on results from \cite{hammache}.  

Thus, by now the primordial abundance of ${}^{6}{\rm Li}$ has been established quite accurately. Taking into account  the latest estimate of the 
${}^{7}{\rm Li}$ abundance ${}^{7}{\rm Li}/{\rm H}= (5.1 \pm 0.4)\times 10^{-10}$  obtained from the most recent data on the ${}^{3}{\rm He}(\alpha,\gamma){}^{7}{\rm Be}$  reaction rate \cite{kontos,bemmerer,confortola}, the resulting isotopic ratio is ${}^{6}{\rm Li}/^{7}{\rm Li} = (1.5 \pm 0.3)\times 10^{-5}$  \cite{LUNA}.  This isotopic ratio is also the result of the present paper. The obtained from the LUNA experiment and indirect ANC  method the Big Bang lithium  isotopic ratio is lower than the previous estimates:   $\,2.3\times 10^{-5}$ \cite{Coc2012}   and $(2- 3.3)\times 10^{-5}$  \cite{coc2014} .
However, invoking the reaction rate following from the present paper  (or from  \cite{muk2011}) and \cite{LUNA} will bring  the result obtained in \cite{Coc2012,coc2014} closer to our and LUNA estimations. 

The established  primordial lithium  isotopic ratio  is by three orders of magnitude lower  then the upper limit determined from the lithium observational 
data in poor-metal, warm dwarf stars what constitutes the second  lithium puzzle. However, the recent publication in Ref. \cite{lind} brings a hope that improving the accuracy of the observational ${}^{6}{\rm Li}$ data can resolve this puzzle without involving non-standard physics.

\section{Summary}

The analysis of the primordial ${}^{2}{\rm H}(\alpha,\gamma){}^{6}{\rm Li}$ reaction is presented. First, the general expression for the  angular distribution of the photons  and specifically for the reaction under consideration is derived. After that the expressions for the total cross sections for the electric dipole and quadrupole transitions are obtained.   The calculated photon's angular distribution, which takes into account the electric dipole and quadrupole transitions and their interference, exhibits the peak at  $\approx 50^{\circ}$.  It provides a recipe for the best experimental kinematics. Note that at the first direct measurements performed by LUNA  [21], the germanium detector was placed at a $90^{\circ}$ angle with respect to the ion beam direction, at which the cross section is significantly smaller than at the peak value.  New measurements with a better geometry can significantly improve the accuracy of the data.
Also the experimental and calculated  $S_{24}(E)$ astrophysical  factors  are presented. Nice agreement between the LUNA data at two Big Bang energies and the potential model calculations based on the ANC  proves the power of the ANC method.

The obtained primordial lithium isotopic ratio in \cite{LUNA} and here  ${}^{6}{\rm Li}/^{7}{\rm Li} = (1.5 \pm 0.3)\times 10^{-5}$ is a very important result in understanding of the second lithium problem. In resolving this puzzle one needs to reconcile both the Big Bang model prediction of the lithium isotopic ratio and the observational data  or to explain their  three orders of magnitude difference. The better the accuracy of the Big Bang ${\rm Li}$ isotopes abundance prediction and the better the agreement with the observational data, the less there will be room for speculations. The results published by LUNA and in this work, ${}^{6}{\rm Li}/^{7}{\rm Li} = (1.5 \pm 0.3)\times 10^{-5}$,  sets up  quite a strong limit on the primordial isotopic ratio from the Big Bang model.  The uncertainty of this ratio is contributed by only $8\%$ 
uncertainty of the ${}^{7}{\rm Li}$ abundance \cite{coc} and by  $22\%$ uncertainty of the ${}^{6}{\rm Li}$ primordial abundance \cite{LUNA}. One of the main conclusions from our work is that the determined optimal kinematics  can significantly improve the accuracy of the ${}^{2}{\rm H}(\alpha,\gamma){}^{6}{\rm Li}$  astrophysical S-factor and, hence, the standard Bing Bang ${}^{6}{\rm Li}$ abundance. 

But even existing predictions of the Big Bang isotopic ratio ${}^{6}{\rm Li}/^{7}{\rm Li} = (1.5 \pm 0.3)\times 10^{-5}$  puts quite a strong upper limit and much more accurate than the observational data. It looks like such a low value of the Big Bang lithium isotopic ratio makes  the second lithium problem even more difficult to resolve.
However, in \cite{lind}  the lithium isotopic analysis in four halo metal-poor stars  was revisited  using, for the first time, a combined 3D and NLTE  modeling technique. This upgraded model systematically reduces the ${\rm Li}$ isotopic ratio in all four analyzed stars 
significantly weakening validity of data  requiring a significant non-standard primordial ${}^{6}{\rm Li}$ production source. 
Hence, it is too early to discuss the compatibility of the Big Bang isotopic ratio  ${}^{6}{\rm Li}/{}^{7}{\rm Li}$, which follows from the latest data on the ${}^{2}{\rm H}(\alpha,\gamma){}^{6}{\rm Li}$ and ${}^{3}{\rm H}(\alpha,\gamma){}^{7}{\rm Li}$  Big Bang reactions, and the observational data of the lithium isotopic ratio in halo, metal-poor, warm stars until the observational analysis will be improved significantly. At least, the work published in Ref. \cite{lind} brings a new hope that the second lithium problem can be resolved without invoking non-standard physics. 

\acknowledgments
A. M. M.  acknowledges that this material is based upon work supported by the U.S. Department of Energy, Office of Science, Office of Nuclear Science, under Award Numbers DE-FG02-93ER40773. It is also supported by the U.S. Department of Energy, National Nuclear Security Administration, under Award Number DE-FG52-09NA29467 and  by the US National Science Foundation under Award PHY-1415656. C.A.B. Acknowledges support from by the U.S. NSF Grant No. 1415656 and the U.S. DOE Grant No. DE-FG02-08ER41533.

\end{document}